\documentclass[nomathfonts,
final,            
numberedheadings 
  ]{aipproc}

\layoutstyle{8x11single}

\newcommand{\beq}{\begin{equation}}
\newcommand{\eeq}{\end{equation}}
  \newcommand{\beql}[1]{\begin{equation}\label{eq:#1}}
  \newcommand{\beqa}{\begin{eqnarray}}
  \newcommand{\eeqa}{\end{eqnarray}}
  \newcommand{\beqas}{\begin{eqnarray*}}
  \newcommand{\eeqas}{\end{eqnarray*}}
  
   \newcommand*{\bA}{\mathbf{A}}

 \newcommand*{\bQ}{\mathbf{Q}}

  \newcommand*{\ep}{\epsilon}
  \newcommand*{\et}{\eta}

  \newcommand*{\ps}{\psi} 
  
  \newcommand*{\si}{\sigma}

  \newcommand*{\Eq}[1]{Eq.~(\ref{eq:#1})}

  \newcommand*{\eq}[1]{(\ref{eq:#1})}

\newcommand*{\bra}[1]{\langle#1|}
\newcommand*{\ket}[1]{|#1\rangle}

\newcommand*{\bracket}[1]{\langle#1\rangle}

\begin{document}

\title[Universal uncertainty principle and 
quantum state control]{Universal uncertainty principle and 
quantum state control under conservation laws}

\author{Masanao Ozawa}{
  address={Graduate School of Information Sciences,
T\^{o}hoku University, Aoba-ku, Sendai,  980-8579, Japan}
}

\begin{abstract}
 Heisenberg's uncertainty principle, exemplified by the $\gamma$ ray 
thought experiment, suggests that any finite precision measurement
disturbs any observables noncommuting with the measured observable.
Here, it is shown that this statement contradicts the limit of the 
accuracy of measurements under conservation laws originally found by 
Wigner in 1950s, 
and should be modified to correctly derive the unavoidable noise
caused by the conservation law induced decoherence. 
The obtained accuracy limit
leads to an interesting conclusion that a widely accepted, but rather naive,
physical encoding of qubits for quantum computing suffers significantly from
the decoherence induced by the angular momentum conservation law.
\end{abstract}

\maketitle

\section{Introduction}

There has been a longstanding confusion on Heisenberg's
uncertainty principle.  
Many text books have incorrectly interpreted 
the mathematical relation between the standard deviations 
as the relation between noise and disturbance exemplified by the
$\gamma$ ray microscope.
In fact, the purported reciprocal relation was not general enough
to hold for all the possible measurements.
Recently, a universally valid noise-disturbance
relation was found by the present author
\cite{03UVR,04URN}
and it has become clear that the new relation plays a
role of the first principle to derive various quantum limits
on quantum measurements and quantum information 
processing in a unified treatment.  
Here, we discuss a consequence of the universal uncertainty
relation for limitations on measurements and quantum state controls
in the presence of nondisturbed quantities 
such as under conservation laws
or superselection rules.

Heisenberg's uncertainty principle, in the naive formulation, 
suggests that
{\em any finite precision measurement disturbs 
any observables noncommuting with the
measured observable.}
Interestingly, the above statement contradicts the limit
on measurements under conservation laws 
known as the Wigner-Araki-Yanase (WAY) theorem,
which allows the finite precision measurement without
disturbing conserved quantities.
Here, we shall resolve the conflict by deriving 
the correct limitation of measurement 
in the presence of nondisturbed quantities.
Then, the obtained formula is shown to quantitatively generalize the
WAY theorem \cite{02CLU}. 
The obtained accuracy limit leads to an interesting conclusion 
that a widely accepted, naive physical encoding of qubits 
suffers from the decoherence induced by  the control system under
the angular momentum
conservation law \cite{02CQC}.  
Various numerical bounds are obtained for inevitable
error probability in physical realizations of 
Hadamard gates.

\section{Universal uncertainty principle}

The uncertainty relation for any observables $A$ and
$B$ is usually formalized by the relation 
\beqa\label{eq:Robertson}
\si (A)\si(B)\ge \frac{1}{2}|\bracket{[A,B]}|
\eeqa
proven first by Robertson \cite{Rob29} in 1929,
where $\si(A)$ and $\si(B)$ are the standard deviations 
of $A$ and $B$ and $\bracket{\cdots}$ stands for the expectation value in
the given state.
For observables $Q$ and $P$ satisfying 
the canonical commutation relation
$QP-PQ=i\hbar$, 
we obtain the uncertainty relation 
\beqa\label{eq:HUR}
\si(Q) \si(P)\ge\frac{\hbar}{2}
\eeqa
proven first by Kennard \cite{Ken27} in 1927. 
The notion of standard deviation in the above formulations depends only on the
state of the system, but does not depend on the measuring
apparatus to be used.
However, it is often explained misleadingly that the physical content of the
above formal relations is that if one
measures observable $Q$, the product of the noise in this
measurement and the disturbance in observable $P$ 
caused by this measurement is no less than  ${\hbar}/{2}$
as claimed by Heisenberg \cite{Hei27} in 1927. 
If we introduce the apparatus dependent quantities, 
the root-mean-square noise $\ep(A)$ in any $A$ measurement
and the root-mean-square disturbance $\et(B)$ of $B$ caused by that measurement,
the above interpretation is expressed by the relation
\beqa\label{eq:HUP}
\ep(A)\et(B)\ge \frac{1}{2}|\bracket{[A,B]}|.
\eeqa
Heisenberg \cite{Hei27} claimed the above interpretation and demonstrated it
by the famous $\gamma$ ray microscope thought experiment.  
However, relation \eq{HUP}, usually called {\em Heisenberg's uncertainty principle},
has been shown not universally valid \cite{02KB5E},
and a universally valid noise-disturbance uncertainty relation, 
the {\em universal uncertainty principle}, has been recently obtained by
the present author \cite{03UVR,04URN} as 
\beqa\label{eq:UUP}
\ep(A)\et(B)+\si(A)\et(B)+\ep(A)\si(B)\ge \frac{1}{2}|\bracket{[A,B]}|,
\eeqa
where the mean and the standard deviations are taken in the state 
just before the measurement.

\section{Uncertainty principle and the Wigner-Araki-Yanase theorem}

The uncertainty principle is a fundamental
source of the noise in measurements, while
the decoherence induced by conservation laws 
is another source of the noise.
Every interaction brings an entanglement in the basis of a conserved
quantity, so that measurements, and any other quantum controls
such as quantum information processing, are subject to the decoherence
induced by conservation laws.
One of the earliest formulations of this fact was given by 
the Wigner-Araki-Yanase (WAY) theorem \cite{Wig52,AY60} 
stating that any observable which does not commute 
with an additively conserved quantity cannot be measured 
with absolute precision.

It is natural to expect that the WAY theorem can be
derived by Heisenberg's uncertainty principle for nondisturbing
measurements.  
Suppose, for instance, that in order to measure the position $Q_{1}$ 
of particle 1 with momentum $P_{1}$, one measures the momentum $P_{2}$  of
particle 2  that has been scattered from particle 1.  
According to the momentum conservation law,
this measurement of $Q_{1}$ does not disturb $P_{1}+P_{2}$.
Thus Heisenberg's uncertainty principle with $\et(P_{1}+P_{2})=0$ 
concludes that we cannot measure $Q_{1}$
even with finite error, i.e., $\ep(Q_{1})=\infty$.

However, the scenario is not that simple.
In reality, we can measure the position $Q_{1}$ 
with finite or even arbitrarily small noise by the above method.
Actually, the WAY theorem does not conclude unmeasurability of any
observables even if they do not commute with the conserved quantity, 
but merely sets the accuracy limit of the measurement with size limited 
apparatus in the presence of bounded conserved quantities. 
In fact, the WAY theorem has a caveat that the noise decreases
if the size of the apparatus increases and that if the apparatus is of
macroscopic size, the noise can be negligible \cite{Yan61,Wig63,Wig76}. 
Thus, the WAY theorem allows an arbitrarily precise measurement with a large
apparatus, but Heisenberg's uncertainty principle 
for nondisturbing measurements does not allow any finite precision
measurement with apparatus of any size.

Now, we abandon Heisenberg's uncertainty principle \eq{HUP}
and consider the problem to find
a correct lower bound for the noise of measurements
in the presence of a nondisturbed quantity.
For this purpose, we return to the universal uncertainty principle \eq{UUP}.
We suppose that the measuring interaction 
does not disturb an observable $B$.
Then, we have $\et(B)=0$, so that by substituting this relation to \Eq{UUP}
we obtain
\beqa\label{eq:UPND}
\ep(A)\sigma(B)\ge\frac{1}{2}|\bracket{[A,B]}|.
\eeqa
The above relation, the {\em universal uncertainty principle for nondisturbing 
measurements},
 represents a correct lower bound for
the noise in measuring observable $A$ using any measuring 
apparatus that does not disturb observable $B$.

The above new formulation of the uncertainty principle 
\Eq{UPND} can be used to derive the quantitative expression of
the WAY theorem as follows.
Suppose that the measuring interaction $U$ satisfies the additive
conservation law 
$[U, L_{1}+L_{2}]=0$,
where $L_{1}$ 
belongs to the object and $L_{2}$ belongs to the apparatus.
Also suppose that the meter observable $M$ in the apparatus 
commutes with the conserved
quantity, i.e., $[M,L_{2}]=0$.
Then, we can conclude that this measurement dose not disturb 
$L_{1}+L_{2}$, thus \Eq{UPND} holds for $B=L_{1}+L_{2}$. 
Since $A$ belongs to the object, 
we have $[A,L_{1}+L_{2}]=[A,L_{1}]$.
Since the object and the apparatus are statistically independent 
before the measurement, we have
$[\si(L_{1}+L_{2})]^{2}=\sigma(L_{1})^{2}+\sigma(L_{2})^{2}$.
Thus, we have derived a quantitative generalization
of the WAY theorem \cite{02CLU,03UPF}
\beqa\label{eq:quantitative-WAY}
\ep(A)^{2}
\ge\frac{|\bracket{[A,L_{1}]}|^{2}}{4\sigma(L_{1})^{2}+4\sigma(L_{2})^{2}},
\eeqa
as a straightforward consequence from \Eq{UPND}.
By the above, the lower bound of the noise decreases with 
the increase of the uncertainty
of the conserved quantity in the apparatus.

\section{Operational Decoherence in Quantum Logic Gates}

In most of current proposals for implementing quantum computing,  
a component of spin of a spin 1/2 system is chosen 
as the computational basis
for the feasibility of initialization and read-out.
For this choice of the computational basis, 
it has been shown \cite{02CQC}
that the angular momentum conservation law limits the accuracy of 
quantum logic operations based on estimating
the unavoidable noise in CNOT gates; see also, Ref.~\cite{Lid03,02CQCReply}.
Here, we shall consider the accuracy of implementing 
Hadamard gates,
which are essential components for quantum Fourier transforms
in Shor's algorithm, and show that Hadamard gates are no easier
to implement under the angular momentum conservation law
than CNOT gates. 

Let $\bQ$ be a spin 1/2 system as a qubit with computational basis
$\{\ket{0},\ket{1}\}$ encoded by 
$S_{z}=(\hbar/2)(\ket{0}\bra{0}-\ket{1}\bra{1})$, 
where $S_{i}$ is the $i$ component of spin for $i=x,y,z$.
Let $H=2^{-1/2}(\ket{0}\bra{0}+\ket{1}\bra{0}+\ket{0}\bra{1}-\ket{1}\bra{1})$
be the Hadamard gate on $\bQ$.
Let $U$ be a physical realization of $H$.
We assume that $U$ is a unitary operator on the composite system
of  $\bQ$ and the ancilla $\bA$ included in the controller of the gate
and that $U$ satisfies the angular momentum conservation law;
see \cite{02CQC,03UPF} for general formulation.
For simplicity, we only assume that the $x$ component of the 
total angular momentum is conserved, i.e, 
$[U,S_{x}+L_{x}]=0$, where $L_{x}$ is the $x$ component 
of the total angular momentum of the ancilla. 

Now, we consider the following process of measuring the operator $S_{z}$ of 
$\bQ$: (i) to operate $U$ on $\bQ+\bA$, and (ii) to measure $S_{x}$ of $\bQ$.
Since $S_{z}=H^{\dagger}S_{x}H$, if $U=H$ the above process would 
measure $S_{z}$ precisely.  
Since each step does not disturb $S_{x}+L_{x}$,
we can apply \Eq{quantitative-WAY} to this measurement and obtain
\beqa
\ep(S_{z})^{2}\ge 
\frac{|\bracket{S_{z},S_{x}}|^{2}}{4\sigma(S_{x})^{2}+4\sigma(L_{x})^{2}}.
\eeqa
In general, the squared-noise $\ep(S_{z})^{2}$ 
amounts to the mean-square error of $U$ from the correct operation of $H$ in the
given input state $\ps$.
Since each error has the squared difference $\hbar^{2}$,
the error probability $P_{e}$ is considered to be 
$P_{e}=\ep(S_{z})^{2}/\hbar^{2}$.
For the input state $\ps=(\ket{0}+i\ket{1})/\sqrt{2}=\ket{S_{y}=\hbar/2}$,
the numerator is maximized as
\beqa\label{eq:bound}
P_{e}=\frac{\ep(S_{z})^{2}}{\hbar^{2}}
\ge \frac{1}{4+4(2\sigma(L_{x})/\hbar)^{2}}.
\eeqa

In the following, we shall interpret the above relation for
bosonic control systems and fermionic control systems separately.
In current proposals, the external electromagnetic field prepared by
laser  beam is considered to be a feasible candidate for the controller
$\bA$ to be coupled with the computational qubits $\bQ$ via the
dipole interaction \cite{NC00}. In this case, the ancilla state 
$\ket{\xi}$ is considered to be a coherent state, for which we have 
$\sigma(N)^{2}=\bracket{\xi|N|\xi}=\bracket{N}$, 
where $N$ is the number operator.  
We assume that the beam propagates to the
$x$-direction with right-hand-circular polarization.  Then, we
have $L_{x}=\hbar N$,  and hence
$(2\sigma(L_{x})/\hbar)^{2}=(2 \sigma(N))^{2}=4\bracket{N}$. 
Thus, from \Eq{bound} we have
\beqa
P_{e}\ge\frac{1}{4+16\bracket{N}}.
\eeqa
Enk and Kimble \cite{EK02} and Gea-Banacloche \cite{Ban02} 
also showed that there is unavoidable error probability in this case
inversely proportional to the average strength of the external field 
by calculations with the model obtained by rotating wave 
approximation.
Here, we have shown the same result only from the angular
momentum conservation law.
If  the field is in a number state $\ket{n}$,
then 
$2\sigma(L_{x}/\hbar)^{2}=2 \sigma(N)^{2}=0$, 
so that we have
\beqa
P_{e}=\frac{\ep(S_{z})^{2}}{\hbar^{2}}\ge\frac{1}{4}.
\eeqa
Thus, if  the field state is a mixture
of number states such as the thermal state, i.e., 
$\si=\sum_{n}p_{n}\ket{n}\bra{n}$, we have also
the lower bound $\ep(S_{z})^{2}/\hbar^{2}\ge{1}/{4}$.
Thus, it seriously matters whether the control field is 
really in a coherent state or a mixture of number states.

We now assume that the ancilla $\bA$ comprises $n$ spin 1/2
systems.
Then, we have
$\sigma(L_{x})\le \|L_{x}\|=\frac{n\hbar}{2}$.
Thus, we have the following lower bound of the gate
error probability
\beqa
P_{e}\ge\frac{1}{4+4n^{2}}.
\eeqa
Thus, it has been proven that if the computational basis is 
represented by the $z$-component of spin, we cannot implement
Hadamard gates within the error probability 
$(4+4n^{2})^{-1}$ with $n$ qubit ancilla by rotationally invariant
interactions such as the Heisenberg exchange interaction.
In the above discussion, we have assumed that the control 
system can be prepared in an entangled state.  However, 
it is also interesting to estimate the error in the case where
we can prepare the control system only in a separable
state.
In this case, we have
$\sigma(L_{x})^{2}\le \sum_{j=1}^{n}\sigma(S_{x}^{(j)})^{2}
\le n\|S_{x}\|^{2}=\frac{n\hbar^{2}}{4}$,
where $ S_{x}^{(j)}$ is the spin component of the $j$th
ancilla qubit so that $L_{x}=\sum_{j=1}^{n}S_{x}^{(j)}$.
Thus, we have the following lower bound of the gate
error probability
\beqa\label{eq:bound2}
P_{e}
\ge \frac{\ep(S_{z})^{2}}{\hbar^{2}}
\ge \frac{1}{4+4n}.
\eeqa
Thus, the error probability is lower bounded by $(4+4n)^{-1}$,
and hence the achievable error can be considered to be
inversely proportional to $4n^{2}$ for entangled control system
but $4n$ for separable control system.  Note that even if
the ancilla is in a separable mixed state, the relation
$(4+4n)^{-1} \le \ep(S_{z})^{2}/\hbar^{2}$
still holds, since $\ep(S_{z})^{2}$ is an affine function
of the ancilla state.

\begin{theacknowledgments}
This work was supported by the SCOPE project
of the MPHPT of Japan,  by the CREST
project of the JST, and by the Grant-in-Aid for Scientific Research of
the JSPS.
\end{theacknowledgments}

\end{document}